\documentclass{article}
\usepackage{spconf,amsmath,graphicx}
\usepackage[colorlinks,linkcolor=blue]{hyperref}
\usepackage{mwe} 
\def\x{{\mathbf x}}

\title{Dual-cycle Constrained Bijective VAE-GAN For Tagged-to-Cine Magnetic Resonance Image Synthesis}
\name{
	\parbox{\linewidth}{\centering Xiaofeng Liu$^1$, Fangxu Xing$^1$, Jerry L. Prince$^2$, Aaron Carass$^2$, Maureen Stone$^3$,\\ Georges El Fakhri$^1$, Jonghye Woo$^1$}
}

\address{$^1$Dept. of Radiology, Massachusetts General Hospital and Harvard Medical School, Boston, MA, USA\\
$^2$Dept. of Electrical and Computer Engineering, Johns Hopkins University, Baltimore, MD, USA\\
$^3$Dept. of Neural and Pain Sciences, University of Maryland School of Dentistry, Baltimore, MD, USA}

\begin{document}
%
\maketitle
\begin{abstract}
Tagged magnetic resonance imaging (MRI) is a widely used imaging technique for measuring tissue deformation in moving organs. Due to tagged MRI's intrinsic low anatomical resolution, another matching set of cine MRI with higher resolution is sometimes acquired in the same scanning session to facilitate tissue segmentation, thus adding extra time and cost. To mitigate this, in this work, we propose a novel dual-cycle constrained bijective VAE-GAN approach to carry out tagged-to-cine MR image synthesis. Our method is based on a variational autoencoder backbone with cycle reconstruction constrained adversarial training to yield accurate and realistic cine MR images given tagged MR images. Our framework has been trained, validated, and tested using 1,768, 416, and 1,560 subject-independent paired slices of tagged and cine MRI from twenty healthy subjects, respectively, demonstrating superior performance over the comparison methods. Our method can potentially be used to reduce the extra acquisition time and cost, while maintaining the same workflow for further motion analyses.
\end{abstract}
\section{Introduction} 

Assessment of an internal organ's deformation has been an important topic in both medical imaging research and clinical practice. One of the most popular modalities to measure internal motion is tagged magnetic resonance (MR) imaging (MRI) with spatially encoded tag patterns~\cite{osman2000imaging}. Tagged MR images with horizontal and vertical tag patterns are acquired as the internal organ, such as the tongue, moves while the tag patterns deform together with the organ; recording all deformation information in the deformed tag patterns. These tag deformations can then be analyzed with motion extraction methods including Harmonic Phase (HARP) based methods, recovering motion fields in either two-dimensional (2D) or three-dimensional (3D) spaces~\cite{xing2017phase, xing2016analysis}. Tagged MRI is an established method for internal organ motion imaging~\cite{petitjean2005assessment} that continues to enjoy widespread use. 

In practice, due to the nature of tagged MRI's encoding method, image resolution on the anatomy of acquired images is relatively low. Although motion fields are computed in each whole MR slice, it is usually necessary to separate the motion field to leave only the part within the organ of interest. As a result, another separate set of dynamic cine MR images without tags are sometimes acquired alongside the tagged MR images as a matching pair~\cite{parthasarathy2007measuring}. Cine MR images have higher resolution than the tagged MR images and are used for segmenting the region of interest, such as the whole tongue \cite{lee2013semi}, in which to compute the motion field. In addition, cine MR images are used to assess surface motion of the organs. However, this extra acquisition can double the time and cost. 

To address this, we systematically investigate a variational autoencoders (VAEs)-based generative model with adversarial training objectives to synthesize cine MRI from tagged MRI as shown in Fig.~\ref{fig:illus}. To the best of our knowledge, this is the first attempt at tagged-to-cine MR image synthesis. Although the L1 loss in VAEs uses pixel-wise pair similarity to provide a good baseline, the results are often blurry \cite{goodfellow2016nips,liu2019hard,liu2020symmetric}. Therefore, an additional generative adversarial network (GAN) \cite{goodfellow2014generative,armanious2020medgan} is further incorporated into our model to enrich the texture, while mitigating potential differences in texture or contrast within real cine MR images. Considering that image translation is largely unconstrained and even ill-posed \cite{zhu2017unpaired}, the anatomical structure of a specific input-output pair is not necessarily consistent if we only align the distributions with GANs. To address this, we introduce an unpaired cycle reconstruction constraint \cite{zhu2017unpaired,liu2017unsupervised} into our VAE-based model so that our proposed dual-cycle constrained bijective VAE-GAN can maintain the anatomical structure.

Both quantitative and qualitative evaluation results using a total of 3,774 paired slices of tagged and cine MRI from twenty healthy subjects show the validity of our proposed dual-cycle constrained bijective VAE-GAN framework and its superiority to conventional VAEs and GANs based image style translation methods.

\section{Methodology}  
Given the paired tagged and cine MR images $\{x^t,x^c\}$, we propose to learn a parameterized mapping $f:x^t\rightarrow \tilde{x}^{t\rightarrow c}$ from the tagged MR images $x^t$ to the generated cine MR images $\tilde{x}^{t\rightarrow c}$ to closely resemble the real cine MR images $x^c$. 

\subsection{VAE-based synthesis}  
A straightforward baseline structure of $f$ can be the VAE, which is constructed with an encoder $Enc$ and a decoder $Dec$. The VAE first maps $x^t$ to a code $z$ in a latent space via the $Enc$, and then decodes $z$ to reconstruct the target image $\tilde{x}^{t\rightarrow c}$ via the $Dec$. For the reconstruction error, we adopt the $L_1$ loss, which can be formulated as $L_{1}({\tilde{x}^{t\rightarrow c}},x^c)= |{\tilde{x}^{t\rightarrow c}}-\x^c|$. The KL-divergence $L_{KL}$ between the latent code $z$ and the prior standard Gaussian is modeled with the standard reparameterization trick \cite{kingma2016improved}. The training objective is to minimize \begin{equation} L_{VAE}=L_{1}({\widetilde{x}^{t\rightarrow c}},x^c)+\alpha L_{KL}(Enc(x^t)||\mathcal{N}(0,I)),\end{equation} where $\alpha$ is a balancing weight for the KL divergence term to penalize the deviation of the distribution of the latent code from the prior distribution. 

A limitation of VAEs, however, is that the generated images tend to be blurry \cite{goodfellow2016nips,liu2019hard} due to the injected noise, limited expressiveness of the inference models, or imperfect element-wise criteria such as the $L_1$ or $L_2$ loss \cite{larsen2015autoencoding}. Although recent studies \cite{kingma2016improved} have greatly improved the predicted log-likelihood, the VAE image generation quality still lags behind GANs \cite{goodfellow2016nips,liu2020auto3d,liu2019feature}. 

In order to enforce perceptual realism with respect to the anatomic structure and improve the quality of the generated textures, we adopt the additional adversarial training procedure \cite{goodfellow2016deep}.

\subsection{Dual-cycle constrained bijective VAE-GAN}  
 The GAN training aims to learn a mapping $f$ with the output $f(x^{t})$ that is indistinguishable from real cine MR images by an adversary trained to classify $f(x^{t})$ apart from $x^c$. A possible solution to combine the objectives of GAN and VAE is following the VAE-GAN framework \cite{larsen2015autoencoding} to configure a discriminator to discriminate real samples from both the reconstructions and the generated examples with sampling Gaussian noise. Similarly, the Pix2Pix \cite{isola2017image} proposes to use the real-fake pair as input to the discriminator to enhance the stability of GANs.  In theory, the GAN's objective can induce an output distribution over $f(x^{t})$ that matches the empirical distribution of $x^c$. The optimal $f$ thereby translates the domain of $x^{t}$ to a domain $f(x^{t})$ distributed identically to $x^c$. 

However, such a translation does not guarantee that the individual input $x^{t}$ and output  $\tilde{x}^{t\rightarrow c}=f(x^{t})$ are paired up in a meaningful way. There are infinitely many mappings $f$ that will induce the same distribution over $f(x^{t})$ \cite{zhu2017unpaired}. While the generated cine MR images may be visually pleasing and realistic, the anatomical shape or structure may not be necessarily consistent with the input tagged MR images. We note that the objective of our tagged-to-cine MR image synthesis is to facilitate the segmentation of the internal organ or to observe surface motion of the organ; therefore, we introduce a stricter requirement with respect to the structural consistency over conventional image translation approaches~\cite{zhu2017unpaired,liu2017unsupervised}.      

\begin{figure}[t]
\begin{center}
\includegraphics[width=1.03\linewidth]{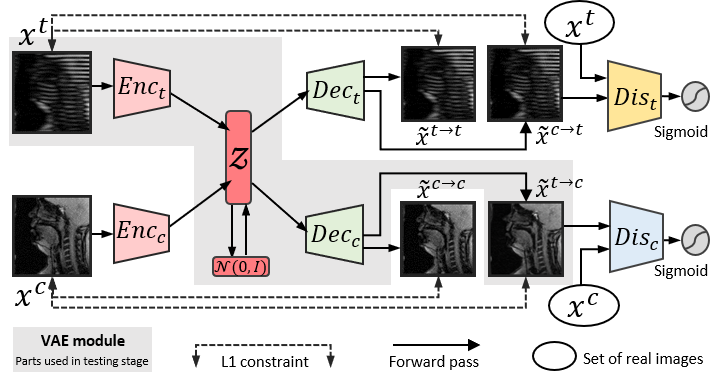}  
\end{center} 
\caption{Illustration of the proposed dual-cycle constrained bijective VAE-GAN. Encoder, decoder, and discriminator have two parallel modules for $x^t$ and $x^c$, respectively. Note that only the gray masked subnets are used for testing.} 
\label{fig:illus}
\end{figure} 

Recently, cycleGAN and its variants \cite{zhu2017unpaired,liu2017unsupervised} have shown outstanding performance in many unpaired image translation tasks. They enforce the generated output with $f:x^t\rightarrow \tilde{x}^{t\rightarrow c}$ can be mapped back to the original image with an additional inverse translator $f^{-1}:x^c\rightarrow \tilde{x}^{c\rightarrow t}$. With two bijective mappings, the distribution of $\tilde{x}^{t\rightarrow c}$ can be largely constrained and therefore the anatomic structure can be better maintained compared with the conventional approaches. Accordingly, we propose to introduce the unpaired cycle constraint \cite{zhu2017unpaired,liu2017unsupervised} for our VAE-based pair-wise synthesis. Our framework, as illustrated in Fig.~\ref{fig:illus}, is based on a dual-branch VAEs and cycle-constrained GANs.

\begin{figure*}[t]
\begin{center}
\includegraphics[width=1.02\linewidth]{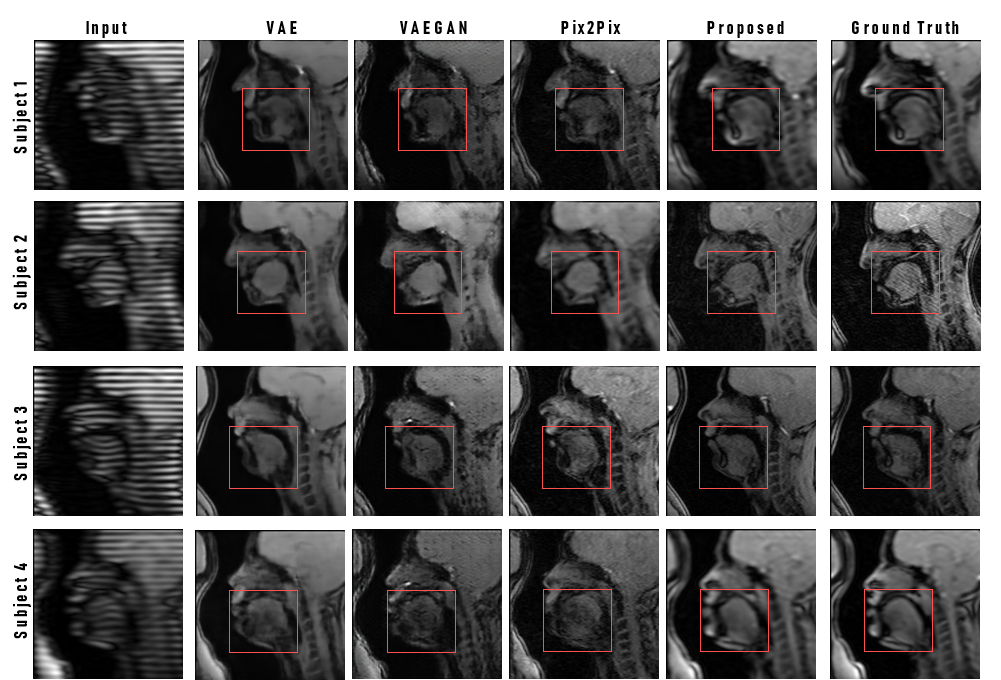} 
\end{center} 
\caption{Comparison of different tagged-to-cine MR generation methods, including the vanilla VAE*, VAE+vanilla GAN~\cite{larsen2015autoencoding}*, Pix2Pix~\cite{isola2017image}*, and our proposed dual-cycle constrained bijective VAE-GAN. * indicates the first attempt at tagged-to-cine MR image synthesis.} 
\label{fig:results}
\end{figure*}

\vspace{+5pt}
\noindent{\bf Detailed structure:} The two encoders (i.e., $Enc_t$ and $Enc_c$) have the same structure and take tagged MR images $x^t$ or cine MR images $x^c$ as input, respectively. The two decoders (i.e., $Dec_t$ and $Dec_c$) mirror the structure of the encoder responsible for tagged or cine MRI generation. The latent space of both tagged and cine MRI is shared \cite{liu2017unsupervised}. 

Similarly, there are two discriminators (i.e., $Dis_t$ and $Dis_c$) for the adversarial training of tagged or cine MRI. For tagged MR images sampled from the tagged MRI domain, $Dis_t$ should output true, while for images generated by $Dec_t$, it should output false.

$Dec_t$ can generate two types of images: 1) images from the reconstruction stream $\tilde{x}^{t\rightarrow t} = Dec_t (z \sim Enc_t(x^t))$ and 2) images from the translation stream $\tilde{x}^{t\rightarrow t} = Dec_t (z \sim Enc_c(x^c))$. Since the reconstruction stream is trained in a supervised manner, it suffices that we only apply the adversarial training to images from the translation stream, $\tilde{x}^{c\rightarrow t}$, given by 
\begin{align}
     &L_1^t(\tilde{x}^{t\rightarrow t},x^t)=|Dec_t(Enc_t(x^t))-x^t|\\
    &L_{Dis_t}= \text{log}(Dis_t(x^t)) + \text{log}(1-Dis_t(\tilde{x}^{c\rightarrow t})).
\end{align}

We apply similar processing to $Dec_c$, where $Dis_c$ is trained to output true for real images sampled from cine MRI and false for images generated from $Dec_c$, given by 
\begin{align}
    &L_1^c(\tilde{x}^{c\rightarrow c},x^c)=|Dec_c(Enc_c(x^c))-x^c|\\
    &L_{Dis_c}= \text{log}(Dis_c(x^c)) + \text{log}(1-Dis_c(\tilde{x}^{t\rightarrow c})).
\end{align}
 

\noindent{\bf Training Strategy:} We jointly optimize the learning objectives of our backbone VAE and the adversarial cycle-reconstruction streams:\begin{align}
    &^{\text{~~~~~~~min}}_{Enc_t,Dec_t}~ L_{VAE}^t+ \beta L_1^t(\tilde{x}^{t\rightarrow t},x^t)+\lambda L_{Dis_t},\\
    &^{\text{~~~~~~~min}}_{Enc_c,Dec_c}~ L_{VAE}^c+ \beta L_1^c(\tilde{x}^{c\rightarrow c},x^t)+\lambda L_{Dis_c},\\
    &~~~~~^{\text{max}}_{Dis_t}~ L_{Dis_t}, ~~~^{\text{max}}_{Dis_c}~ L_{Dis_c},
\end{align} where $L_{VAE}^t=L_{1}({\widetilde{x}^{t\rightarrow c}},x^c)+\alpha L_{KL}(Enc_t(x^t)||\mathcal{N}(0,I))$, and $L_{VAE}^c=L_{1}({\widetilde{x}^{c\rightarrow t}},x^t)+\alpha L_{KL}(Enc_c(x^c)||\mathcal{N}(0,I))$ are the backbone VAE objectives for $x^t$ and $x^c$, respectively. The hyper-parameters $\beta$ and $\lambda$ control the weights of the cycle-based reconstruction objective term and the discriminate term, respectively. Inheriting from GAN, the $Enc$-$Dec$ and $Dis$ play the round-based min-max adversarial game to improve each other to find a saddle point. In practice, we usually sample a similar number of $x^t$ and $x^c$ in training \cite{liu2017unsupervised}.  

\vspace{+5pt}
\noindent{\bf Testing Tagged-to-Cine Translation:} Targeting our tagged-to-cine MR synthesis objective, after training, we can obtain the translation functions by assembling a subset of the subnetworks, i.e., $Enc_t$ and $Dec_c$. Therefore, given tagged MR images $x^t$, we can generate its corresponding $\tilde{x^{t\rightarrow c}}=Dec_c(Enc_t(x^t))$. The involved subnets are indicated with the gray mask in Fig.~\ref{fig:illus}.
 
\section{EXPERIMENTS AND RESULTS} 
 
\subsection{Data Acquisition and Subjects} 

For the experiments performed in this study, a total of 3,774 paired tagged and cine MR images from twenty healthy subjects were acquired while speaking an utterance, ``a souk” in line with a periodic metronome-like sound. MRI scanning was performed on a Siemens 3.0T TIM Trio system with a 12-channel head coil and a 4-channel neck coil using a segmented gradient echo sequence~\cite{xing2016analysis}. The field of view was 240$\times$240 mm with an in-plane resolution of 1.87$\times$1.87 mm and a slice thickness of 6 mm. The image sequence was obtained at the rate of 26 fps. Both cine and tagged MRI are in the same spatiotemporal coordinate space.

\subsection{Experimental Setup} 

We used 10 subjects (1,768 slice pairs) for training, 2 subjects (416 slice pairs) for hyper-parameter validation, and 8 subjects (1,560 slice pairs) for evaluation. Tagged MR images with horizontal tag patterns were used in our evaluation. For fair comparison, we resized the tagged and cine MR images to 256$\times$256 and adopted the $Enc$, $Dec$, and $Dis$ backbones from \cite{isola2017image} for all of the methods.

Our framework was implemented using the PyTorch deep learning toolbox. The training was performed on an NVIDIA V100 GPU, which took about 4 hours. In practice, translating one tagged MR image to a cine MR image in testing with only $Enc_t$ and $Dec_c$ took about 0.1 seconds. 

In order to align the absolute value of each loss, we set different weights for each part with grid searching on the validation data. Specifically, we set weight $\alpha=1$, $\beta=1$, and $\lambda=0.5$. We used Adam optimizer for training. The learning rate was set at $lr_{Enc,Dec}=1\mathrm{e}{-3}$ and $lr_{Dis}=1\mathrm{e}{-4}$ and the momentum was set at 0.5.

\subsection{Qualitative Evaluations} 

The synthesis results using VAE \cite{kingma2016improved}, VAE-GAN \cite{larsen2015autoencoding}, Pix2Pix \cite{isola2017image}, and our proposed method are shown in Fig.~\ref{fig:results}. The proposed framework successfully synthesized the cine MR images, which are consistent with the target ground truth cine MR images. The VAE only model simply relies on the pixel-wise matching, thus resulting in relatively blurry outputs. There is a relatively large difference in contrast and texture within the tongue region between the real cine MR images, which could affect subsequent analyses. Thus, the added-on adversarial objective is expected to enrich the fine-grained tissue texture \cite{goodfellow2016deep}. VAE-GAN \cite{larsen2015autoencoding} adopts the vanilla GAN, which is difficult to optimize, and often leads to the well-known problem of mode collapse. By using the pair-wise inputs to the discriminator, the model collapse in Pix2Pix \cite{isola2017image} was relatively alleviated. Although the generative outputs were visually realistic, the unconstrained image translation yielded distortion in the anatomical structure. In contrast, our proposed dual-cycle constrained bijective VAE-GAN with the cycle reconstruction constraint generated visually pleasing results with better structural consistency, which is important for subsequent analyses.

\begin{table}[t]
\centering
\caption{Numerical comparisons of four methods in testing across 1,560 slice pairs. The best and the second best results are \textbf{bold} and \underline{underlined}, respectively.} \vspace{+5pt}
\resizebox{1\linewidth}{!}{
\begin{tabular}{c|c|c|c|cccc|ccc}
\hline
Methods& L1~$\downarrow$SSIM~$\uparrow$ &	PSNR~$\uparrow$ 	&  IS~$\uparrow$\\\hline\hline
VAEonly                             	&\textbf{148.8}$\pm$0.1&	0.9486$\pm$0.0014&	32.68$\pm$0.06&	11.13$\pm$0.11\\
VAE-GAN \cite{larsen2015autoencoding}	&151.4$\pm$0.2&	0.9507$\pm$0.0013&	34.14$\pm$0.05&	11.68$\pm$0.17\\
Pix2Pix \cite{isola2017image}       	&150.2$\pm$0.2&	\underline{0.9612}$\pm$0.0011&	\underline{36.81}$\pm$0.07&	\underline{13.77}$\pm$0.15\\\hline
Proposed                                &\underline{149.6}$\pm$0.2&	\textbf{0.9746}$\pm$0.0015& \textbf{38.72}$\pm$0.07&  \textbf{15.58}$\pm$0.12\\\hline
\end{tabular}
}
\label{tabel:1}
\end{table}

\subsection{Quantitative Evaluations}  
The synthesized images were expected to have realistic-looking textures, and to be structurally coherent with its corresponding ground-truth $x^c$. For quantitative evaluation, we adopted widely used evaluation metrics: mean L1 error,  structural similarity index measure (SSIM), peak signal-to-noise ratio (PSNR), and inception score (IS) \cite{song2018contextual}. 

Table \ref{tabel:1} lists numerical comparisons between the proposed framework, VAE \cite{kingma2016improved}, VAE-GAN \cite{larsen2015autoencoding}, and Pix2Pix \cite{isola2017image} for the 8 testing subjects.~The proposed dual-cycle constrained bijective VAE-GAN outperformed the other three comparison methods with respect to SSIM, PSNR, and IS. We note that all of the compared methods have the L1 minimization objective.
 
\section{Conclusion and Future Work}  

We presented a novel framework to synthesize cine MRI from its paired tagged MRI. We systematically investigated VAE with the GAN objectives, and proposed a novel dual-cycle constrained bijective VAE-GAN. The adversarial training introduced the realistic texture, while mitigating potential differences in contrast and texture within the real cine MR images. In addition, importantly, the qualitative evaluation shows the anatomical structure of the tongue was better maintained with the additional cycle reconstruction constraint. Our experimental results demonstrated that our method surpassed the comparison methods as quantitatively and qualitatively assessed. The synthesized cine MRI can potentially be used for further segmenting the tongue and observing surface motion, and, in our future work, we will develop a segmentation network in conjunction with the proposed network using the synthesized cine MR images.

\section{COMPLIANCE WITH ETHICAL STANDARDS}\vspace{-5pt}
The subjects signed an informed consent form, and data were collected in accordance with the protocol approved by the Institutional Review Board (IRB) of the University of Maryland, Baltimore.~The retrospective use of the data was granted by IRB of Massachusetts General Hospital. 
\section{ACKNOWLEDGMENTS}\vspace{-5pt}
{This work is supported by NIH R01DC014717, R01DC018511, R01CA133015, and P41EB022544.}
\bibliographystyle{IEEEbib}
\bibliography{ref5page}

\end{document}